\documentclass[apj]{emulateapj}

\newcommand{\grad}{^{\mbox{\small o}}}

\newcommand{\cm}{\mbox{~cm}}
\newcommand{\kpc}{\mbox{~kpc}}
\newcommand{\pc}{\mbox{~pc}}

\newcommand{\kms}{\mbox{~km s}^{-1}}

\begin{document}

\title{3D MHD Modeling of the Gaseous Structure of the Galaxy:
  Synthetic Observations.}

\author{Gilberto C. G\'omez\altaffilmark{1}}
\affil{Department of Astronomy - University of Wisconsin,
  475 N. Charter St., Madison, WI 53706 USA}
\email{gomez@wisp.physics.wisc.edu}
\and
\author{Donald P. Cox}
\affil{Department of Physics - University of Wisconsin,
  1150 University Ave., Madison, WI 53706 USA}
\email{cox@wisp.physics.wisc.edu}

\altaffiltext{1}{Now at Department of Astronomy -
  University of Maryland, College Park, MD 20742;
  e-mail: {\tt gomez@astro.umd.edu}}

\begin{abstract}

We generated synthetic observations from the four-arm model
presented in \citet{gom04} for the Galactic ISM in the
presence of a spiral gravitational perturbation.
We found that velocity crowding and diffusion have a strong
effect in the $l-v$ diagram.
The $v-b$ diagram presents structures
at the expected spiral arm velocities, that can be explained by
the off-the-plane structure of the arms
presented in previous papers of this series.
Such structures are observed in the Leiden/Dwingeloo
\ion{H}{1} survey.
The rotation curve, as measured from the inside of the modeled
galaxy, shows similarities with the observed one for
the Milky Way Galaxy, although it has large
deviations from the smooth circular rotation
corresponding to the background potential.
The magnetic field inferred from a synthetic synchrotron map
shows a largely circular structure, but with interesting deviations
in the midplane due to distortion of the field from circularity
in the interarm regions.

\end{abstract}

\keywords{ISM: kinematics and dynamics --- MHD
      --- galaxies: spiral, structure}


\section{Introduction.}

\begin{figure}[!b]
\plotone{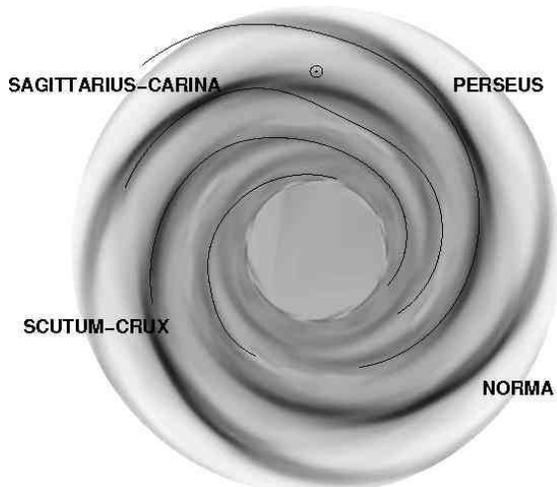}
\caption{
Surface density of the simulation, compared
with the Milky Way's spiral arms, as traced by \citet{geo76},
modified by \citet{tay93}.
The position chosen for the observer in the following synthetic maps
is also presented.
The galactocentric distance for the Sun was chosen to be
$8 \kpc$.
}
\label{arms}
\end{figure}

\begin{figure}[!b]
\plotone{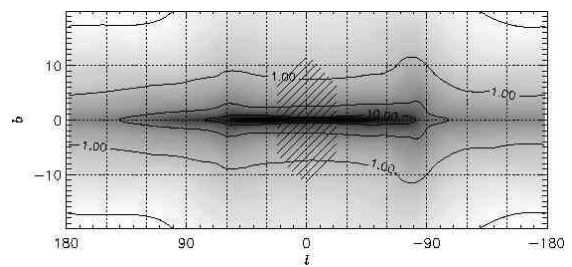}
\caption{
Column density of the gas, for a region around the galactic
plane.
The contours are in geometric sequence and labeled
in units of $\kpc \cm^{-3}$.
The shaded region marks the inner limit of the simulation domain.
}
\label{all_sky}
\end{figure}

Our position inside the Milky Way Galaxy allows us to make
observations at a much higher spatial resolution that we could
do in other disk galaxies.
But that same fact makes it much more difficult to infer the
large scale characteristics of our home galaxy.
A lot of the current questions of the spiral structure of the
Milky Way could be resolved if we knew the position and full
velocity vector of the observed gas.
Numerical studies of large scale galactic structure have proved
to be very valuable in discerning the sought after characteristics.
Nevertheless, this is a very complicated problem and,
so far, it is impossible to include all the physics involved.
Therefore, modelers must decide which parts of the problem are
not going to be considered, in the hope that those neglected
will have little influence in the overall conclusions.
The models presented here
do not include self-gravity of the gas,
supernova explosions or other energetic events, and
have uncomfortably low spatial resolution.
They include a substantial
magnetic field, a high thermal pressure (to represent tangled
fields, cosmic rays, and subgrid turbulence),
and the extra degree of freedom of three
dimensions.
We believe these are definitive factors that have not
been sufficiently explored.
The thermal pressure was also adjusted to drop sharply at high
densities to encourage the formation of denser structures.

\begin{figure*}
\plotone{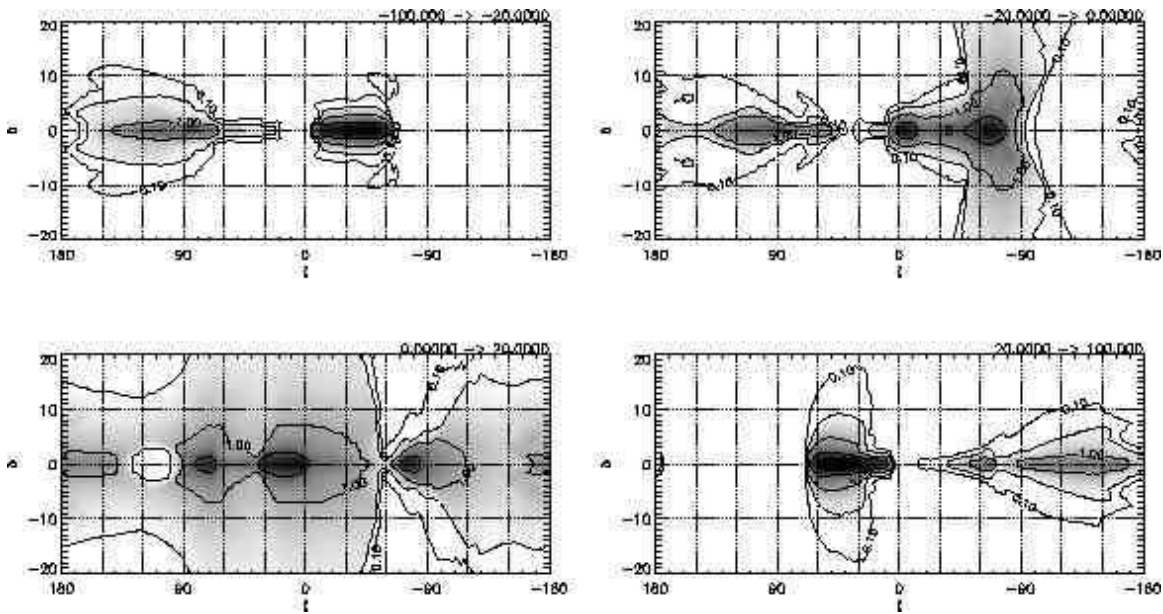}
\caption{
Same as Figure \ref{all_sky}, but restricted in the
line-of-sight velocity component of the gas.
The range of velocity integration is presented in the top-right
corner of each panel, in units of $\kms$.
Most of the higher column density elements in these plots can
be traced to spiral arms (or superposition of arms, see text
for a discussion).
But the one at about $l=75\grad$ in the $0 \kms<v_{los}<20 \kms$
is in an interarm direction, in which the streaming motions
generate a small range in line-of-sight velocities over a large
path length.
}
\label{all_sky_vel}
\end{figure*}

In the first two papers of this series
\citep[Papers I and II from here on]{gom02,gom04},
we presented the results
of our simulations of the ISM response to a spiral gravitational
perturbation.
We showed that the extra stiffness that the magnetic field adds
to the gas makes it develop a combination of a shock and a
hydraulic jump with significant complications added by
vertical bouncing.
This jump/shock leans upstream above the plane (more in the two-arm
models than the four-arm ones),
ahead of the main
gas concentration in the midplane.
As it shocks, the gas shoots up to higher $z$ in a way similar to
water jumping over an obstacle in a riverbed.
The gas then accelerates as it runs over the arm, and falls
down behind it, generating a secondary set of shocks.
In  the two-arm cases, the gas bounces back up, generating interarm
structures that mimic the ones found at the arms.

In this work, we'll focus in the four-arm case, designed to have
spiral arms similar to those traced by \citet{geo76},
as modified by \citet{tay93}.
In Figure \ref{arms} we show
the surface density of the four-arm model from Paper II,
along with the aforementioned arm pattern, and the corresponding
position of the Sun.
Notice that the scale of the spiral arms has been reduced
so that the distance from the Sun to the galactic center
is $8\kpc$, as in our model.
In Section \ref{all_sky_sec} we present all sky column density
maps in radial velocity ranges;
in Section \ref{lv_sec} we present synthetic longitude-velocity
diagrams;
in Section \ref{vb_sec} we present velocity-latitude diagrams,
which we believe have a definite signature of these models;
in Section \ref{rotcurve_sec} we present the rotation
curve that would be measured in this galaxy
as affected by the spiral arms;
in Section \ref{distances_sec} we analyze its effect on the
measured kinematic distances in the galactic plane;
in Section \ref{corotation_sec} we examine the rotation
of gas above the midplane;
in Section \ref{synch_sec} we present an all sky synchrotron
map;
and in Section \ref{conclusions_sec} we present our conclusions.


\section{All sky maps.}\label{all_sky_sec}

\begin{figure}
\plotone{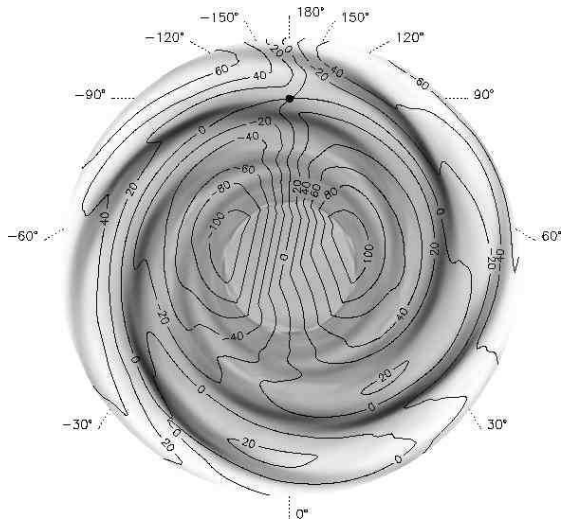}
\caption{
Surface density of the simulation with
the midplane line-of-sight velo\-ci\-ties overplotted.
The position of the observer is marked by the dot.
Galactic longitudes are also labeled.
}
\label{radial_vel}
\end{figure}

Figure \ref{all_sky} shows a map of the integrated column density
of the simulation, as seen from the position of the observer
marked in Figure \ref{arms}, in galactic coordinates.
The grayscale shows the column density, with
contours in a geometric sequence
and labeling in units of $\kpc \cm^{-3}$.
(The reader should keep in mind that our model spans only
from $3 \kpc$ through $11 \kpc$ in radius,
and up to $1 \kpc$ in $z$.
The shaded region in the galactic center direction shows
the angular extent of the central ``hole'' in our simulation grid.
In addition, the full strength of the perturbation
is applied only for $r > 5\kpc$, and therefore, the useful part
of the grid extends from $r=5$ to $11\kpc$.)
Two vertical protuberances are clear in this Figure,
corresponding to the
directions tangent to the Sagittarius arm, at
$l \sim 60 \grad$ and $l \sim -75$.
Imprints corresponding to other arms are also present;
they are harder to pick up in this Figure, but become evident
when we restrict the line-of-sight integration to certain
velocity ranges, as in Figure \ref{all_sky_vel}. (A map with
the line-of-sight component of the velocities for the midplane
is presented in Figure \ref{radial_vel}).
The Perseus arm appears in all the velocity ranges, but it
is more prominent in $l > 90 \grad$ at negative velocities,
and $l< -90 \grad$ at $v>20 \kms$.
Also prominent are: a superposition of the Perseus, Scutum
and Norma arms at between $l \sim 0\grad$ and $-60\grad$
at $v<-20 \kms$;
the Sagittarius arm from $l \sim -60\grad$ to $-90\grad$
at intermediate
negative velocities, with a very large vertical extension;
and a superposition of the Sagittarius and Scutum arms
at from $l \sim 0\grad$ to $60\grad$ for large positive velocities.

The diagram for $0 < v_{los} < 20 \kms$ features
three large column density elements.
The one at $l \sim -75\grad$ corresponds to the Sagittarius arm.
At $l \sim 0\grad$ to $30\grad$, we see a narrow stripe of the
whole inner galaxy, including
the Perseus, Scutum, and Sagittarius arms.
The third element, at $l \sim 75\grad$,
corresponds to a tangent direction through the region
between the Sagittarius and Perseus arms, in which lower density
gas spans a large distance with a small range of velocities.



\begin{figure}[b]
\plotone{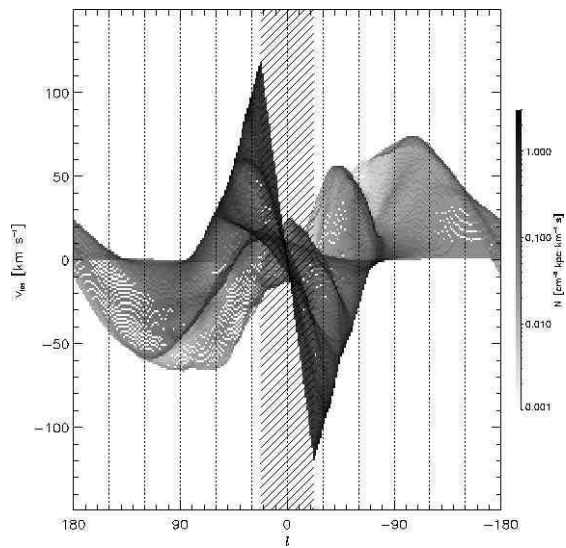}
\caption{
$l-v$ diagram for $b=0\grad$. The shaded region marks
the ``hole'' in our simulation grid.
Many features of the observed $l-v$ diagrams are reproduced:
the asymmetry in the inner rotation curve, the non-zero mean
velocity in the $l=180\grad$ direction, and the high column
density ridges, usually associated with spiral arms.
}
\label{lv_4}
\end{figure}

\begin{figure}[t]
\plotone{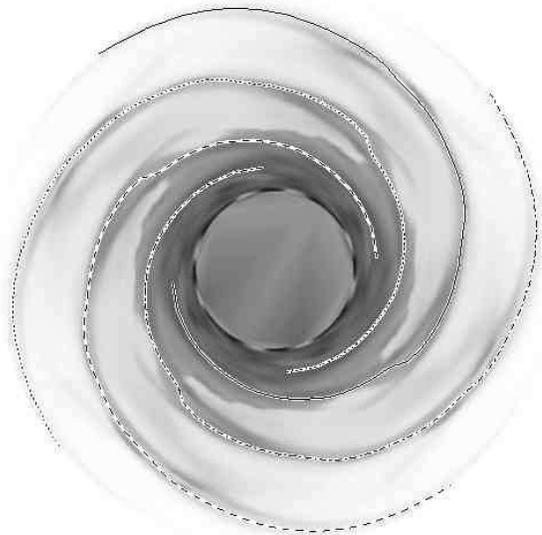}
\caption{
Loci of the spiral arms traced by fitting a sinusoidal
function to the vertical column density, for each radius.
The solid line traces the Perseus arm, dashed-dotted the Norma arm,
dashed the Scutum, and dotted the Sagittarius arm.
The grayscale is the midplane density of the model.
}
\label{arms_locus}
\end{figure}

\begin{figure}[!b]
\plotone{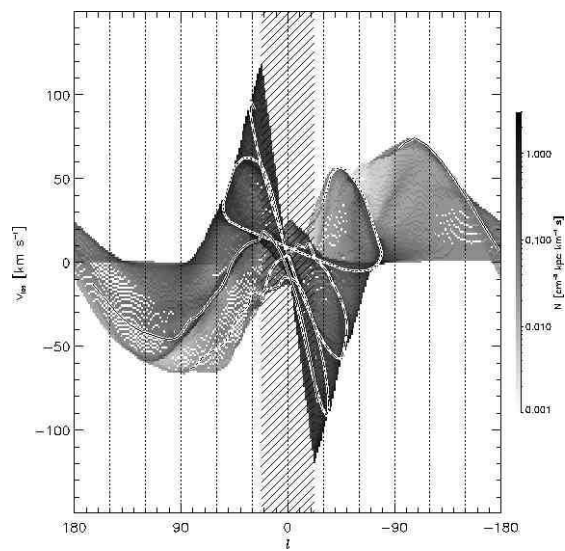}
\caption{
Same as in Figure \ref{lv_4}, with the spiral arms from
Figure \ref{arms_locus} traced on it.
The solid line traces the Perseus arm, dashed-dotted the Norma arm,
dashed the Scutum, and dotted the Sagittarius arm.
Notice that not all the ridges correspond to arms, and not all the
arms trace back to ridges.
}
\label{lv_arms}
\end{figure}

\section{$l-v$ diagram.}\label{lv_sec}

Longitude-velocity diagrams are a straightforward way of
presenting the data-cubes, and it is relatively easy to extract
information about global properties of the Galaxy.
Figure \ref{lv_4} shows a longitude-velocity diagram, as
seen by an observer situated as marked in Figure \ref{arms},
at galactic latitude $b=0\grad$.
Again, the shaded region marks the region around the galactic
center that we do not include in our model.
Several features can be pointed out.
The rotation curve, as traced by the extremum velocity of the
gas in the inner galaxy, has small deviations
from a flat curve which are not symmetric about $l=0\grad$
(see Section \ref{distances_sec}).
Even without velocity dispersion intrinsic to the gas in the model,
there is a fair amount of gas moving at forbidden velocities
in the general direction of the galactic anticenter
(positive in the second quadrant, negative in the third).
Also, the gas in the anti-center direction has a positive
mean velocity,
while the envelope of the emission averages to zero
at $l \sim 170 \grad$.
These characteristics depend strongly on the chosen position
for the observer.
Features similar to these are observed
in $l-v$ diagrams from Milky Way's \ion{H}{1} surveys,
although details
(like the longitude of the zero velocity around the anticenter)
do not necessarily coincide with our model.
The proximity of our outer boundary ($3 \kpc$ in the
anticenter direction) might have some influence on this.

Ridges and intensity enhancements in this diagram are
usually interpreted as spiral arms. 
Since in our model we have the advantage of knowing exactly where
the material
is and with which velocity it is moving, we can trace the
gaseous spiral arms into the simulated $l-v$ diagram.
We found the position of the spiral arms by fitting,
for each radius in the simulation grid,
a sinusoidal function along azimuth
to the vertical column density of the gas.
Figure \ref{arms_locus} presents the result of the fit, while
Figure \ref{lv_arms} traces the spiral arms into the $l-v$ diagram.
Most of the ridges in this diagram correspond
to spiral arms, although the relation is not one-to-one.
For example, around $l = 120\grad$, at the Perseus arm,
the line of sight goes through a large velocity gradient,
which spreads the arm in velocity, and diffuses the ridge.
The converse also happens: lower intensity ridges that are
not related to spiral arms are generated when the velocity
gradient is small, and large spatial extents condense into
a small velocity range,
for example, at $(l,v) \sim (-90\grad, 30 \kms)$.
The capacity of the velocity field to create or destroy structures
in this diagram with little regard of the underlying gas density has
been long known \citep{bur71, mul86}.



\section{$v-b$ diagram.}\label{vb_sec}

Another natural way of presenting data cubes is the
velocity-latitude diagram.
For our model, the $v-b$ diagram shows the signature
of the vertical structure of the spiral arms.

\begin{figure}[b]
\plotone{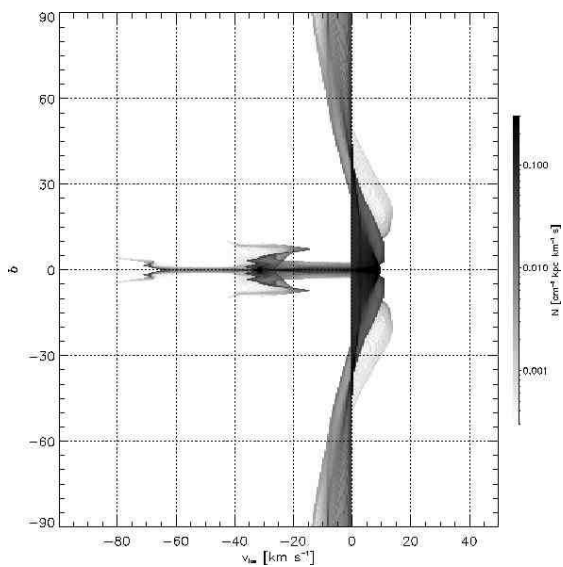}
\caption{
$v-b$ diagram along $l=75\grad$.
Notice the gas at negative velocities around the galactic pole
directions and the vertical ridge at $v \sim -10\kms$ above
$b\sim 50\grad$, created by velocity crowding.
The structure is truncated for $v < -70 \kms$ by encounter
with the edge of the simulation grid.
}
\label{latvel_75}
\end{figure}

Figure \ref{latvel_75} shows the $v-b$ diagram for the $l=75 \grad$
direction.
The position chosen for the observer places it just downstream
from the Sagittarius arm, where the gas is falling down.
Therefore, Figure \ref{latvel_75} shows gas with negative
velocities at large galactic latitudes.
This is consistent with the observations by \citet{die64}, the
WHAM project \citep{haf03}, and other authors,
which found that gas around the
galactic poles has a mean negative velocity.
Notice the higher intensity ridge that runs from $v=0\kms$
at $b \sim 40\grad$ to
about $v=-8 \kms$ at the galactic pole.
That ridge is generated by crowding of the falling gas in
velocity space.
A spectrum taken toward those latitudes would show a line that could
be interpreted as a cloud, although the gas has no spatial
concentration.

\begin{figure}[t]
\plotone{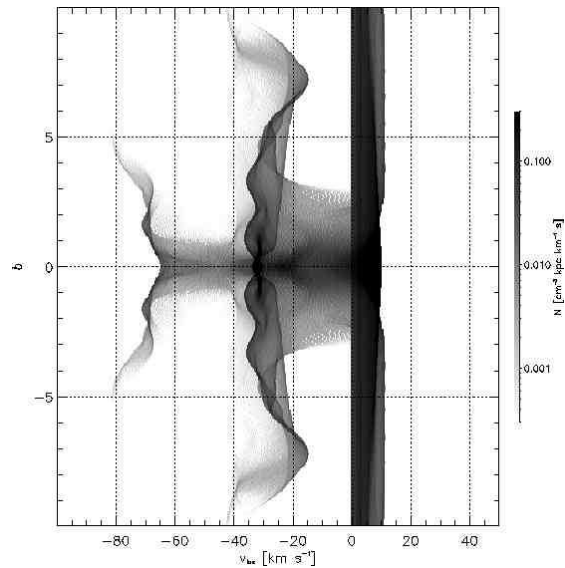}
\caption{
Same as Figure \ref{latvel_75}, zooming to $20\grad$
around the galactic plane.
The horizontal mushroom shapes are 
consequences of the three dimensional
structure of the spiral arms, as discussed in the text.
For this longitude, the structure at $v \sim -30 \kms$ corresponds
to the Perseus arm, and the structure at $v \sim -70 \kms$
corresponds to the Norma arm.
See also Figure \ref{dist_lat_75}.
}
\label{latvel_75_zoom}
\end{figure}

Figure \ref{latvel_75_zoom} zooms into a $20\grad$ region around the
galactic midplane.
Examination of Figure \ref{radial_vel} shows that
such line-of-sight crosses the Perseus arm at
$v \sim -30\kms$ and the Norma arm at $v \sim -70\kms$,
very close to the simulation edge.
At those approximate velocities,
Figures \ref{latvel_75} and \ref{latvel_75_zoom},
show ``mushroom'' shaped structures, with a
relatively narrow, horizontal stem and large
vertical cap on the left edge of the stem.
These are the characteristic signatures of the
vertical structure of the gaseous arms, along with a tendency
of the tip of the cap to bend slightly back over the stem,
to less negative velocities at higher latitude.

\begin{figure}
\plotone{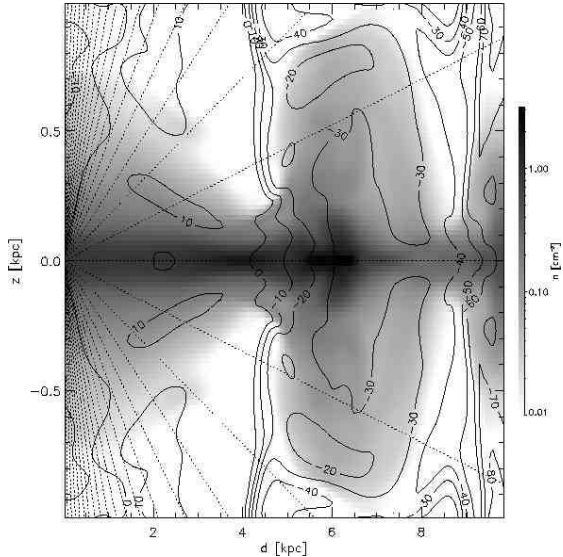}
\caption{
Logarithm of density (grayscale)
and line-of-sight component of the velocity
(contours) along the $l=75\grad$ direction.
The velocity contours are labeled in $\kms$ units.
The dotted lines mark the latitude directions every $5 \grad$.
The abscissa is the distance from the position
of the observer.
As the gas falls from the Sagittarius arm, it compresses the disk
before flowing into the Perseus arm (at $d \sim 6 \kpc$), where it
is slowed down and forms the narrow stem
of the ``mushroom'' structure in Figure
\ref{latvel_75_zoom}. The gas swells around the spiral arm, and the
high latitude flow speeds up as it passes over the midplane lump.
This flow shows in Figure \ref{latvel_75_zoom} as the tips of the mushroom cap
bent toward less negative velocities. After the arm, the gas falls
into the midplane, compressing the disk before the flow shocks into
the Norma arm (near the outer boundary).
}
\label{dist_lat_75}
\end{figure}

In order to guide the following discussion, we present in Figure
\ref{dist_lat_75} the density (grayscale) and line-of-sight
component of the velocity (contours) along a vertical plane in the
$l=75\grad$ direction.
Since we are looking at the arm from the concave side, the gas
moves through it from left to right, though not parallel
to the plane of the Figure.
The reader should keep in mind that these velocities are the result
of the presence of the arms on top of the galactic rotation.
If the gas were in purely circular orbits, we should see positive
velocities up to the solar circle, at a distance of $4.1 \kpc$,
and then negative
velocities, monotonically decreasing until the edge of the grid.
This general pattern is found in the Figure, with velocities
increasing from 0 to $+10\kms$ at $\sim 2\kpc$, then back to zero
at $\sim 4\kpc$, going increasingly negative, though not
monotonically, beyond that.

Let us concentrate on the Perseus arm, at about $6\kpc$.
Just upstream from the arm,
there is a vertically thin distribution of material formed by the
downflow from the Sagittarius arm.
The encounter of this material with the
arm decelerates the gas,
appearing as a rapid succession of
decreasing velocity contours at $d \sim 4.5\kpc$.
That velocity gradient spreads the vertically thin gas
structure along the horizontal axis of
the $v-b$ diagram, creating the stem structure in the
midplane.
Beyond $\sim 5\kpc$, we have  the vertically swelled structure
of the arm itself, which appears
in Figure \ref{latvel_75_zoom} as the vertically extended mushroom
cap.
Between $b \sim 5\grad$ and $7\grad$,
the gas is speeding up above the arm
increasing its line-of-sight velocity, causing the tip of the cap to
slightly bend back over the stem.
(The reversal of this trend between $b \sim 7.5\grad$ and $10\grad$
involves a very small amount of material very close to the vertical
boundary and could be an artifact.)
Within the arm, the radial velocity has one or more extremes,
creating caustics as the gas doubles back in velocity space.
This behavior repeats as we approach the Norma arm, but
we reach the simulation boundary before developing the full arm
structure, and we get only the stem and the beginning
of the cap.

When we restricted the integration to the thin preshock
interarm region, we noticed that
its $v-b$ imprint is very small.
But, as seen in Figure \ref{dist_lat_75},
the true ``interarm'' region is much broader and includes
the high $z$ part of the cap of the previous arm,
after the midplane density has decreased.


\begin{figure*}
\plottwo{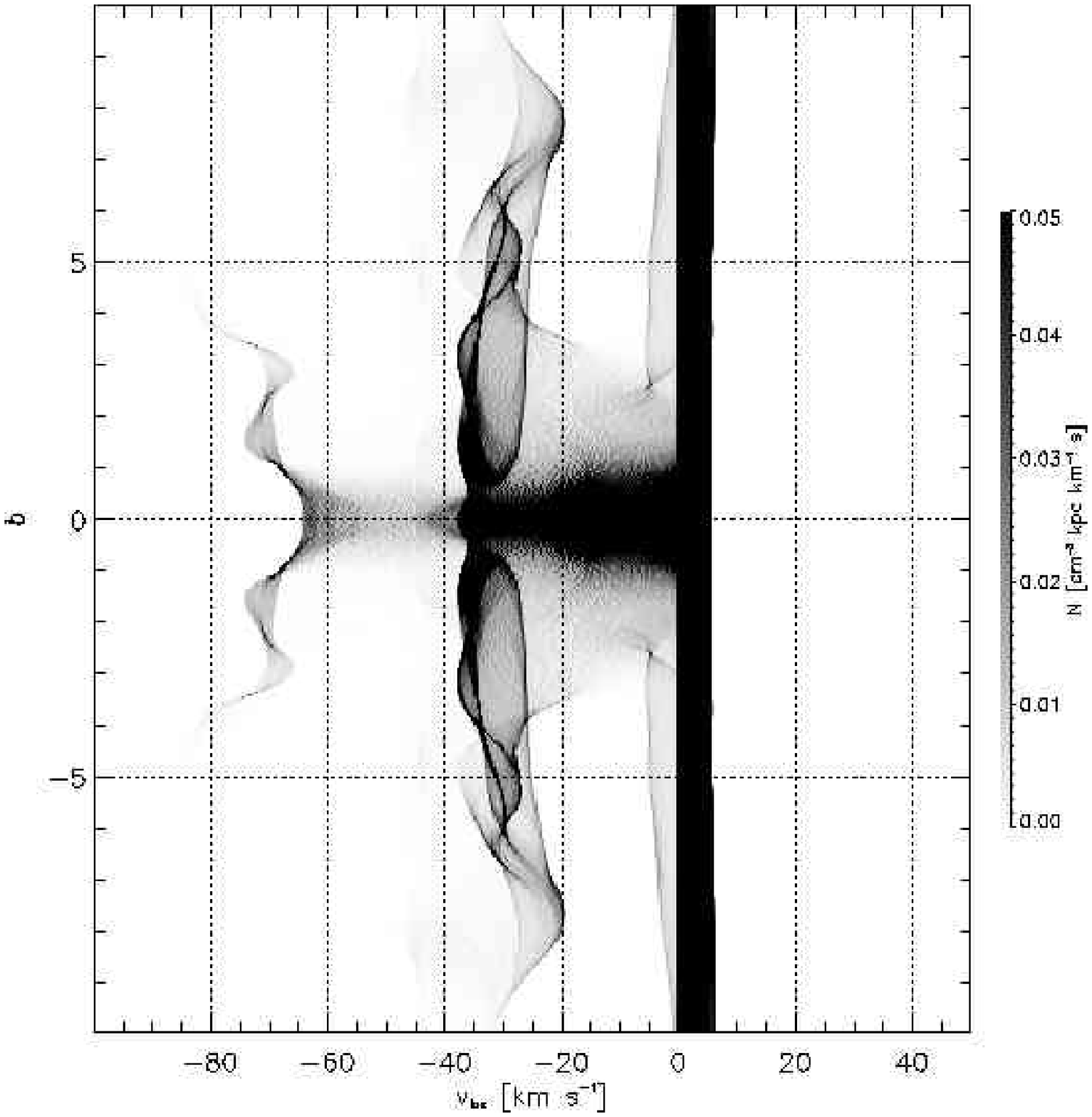}
        {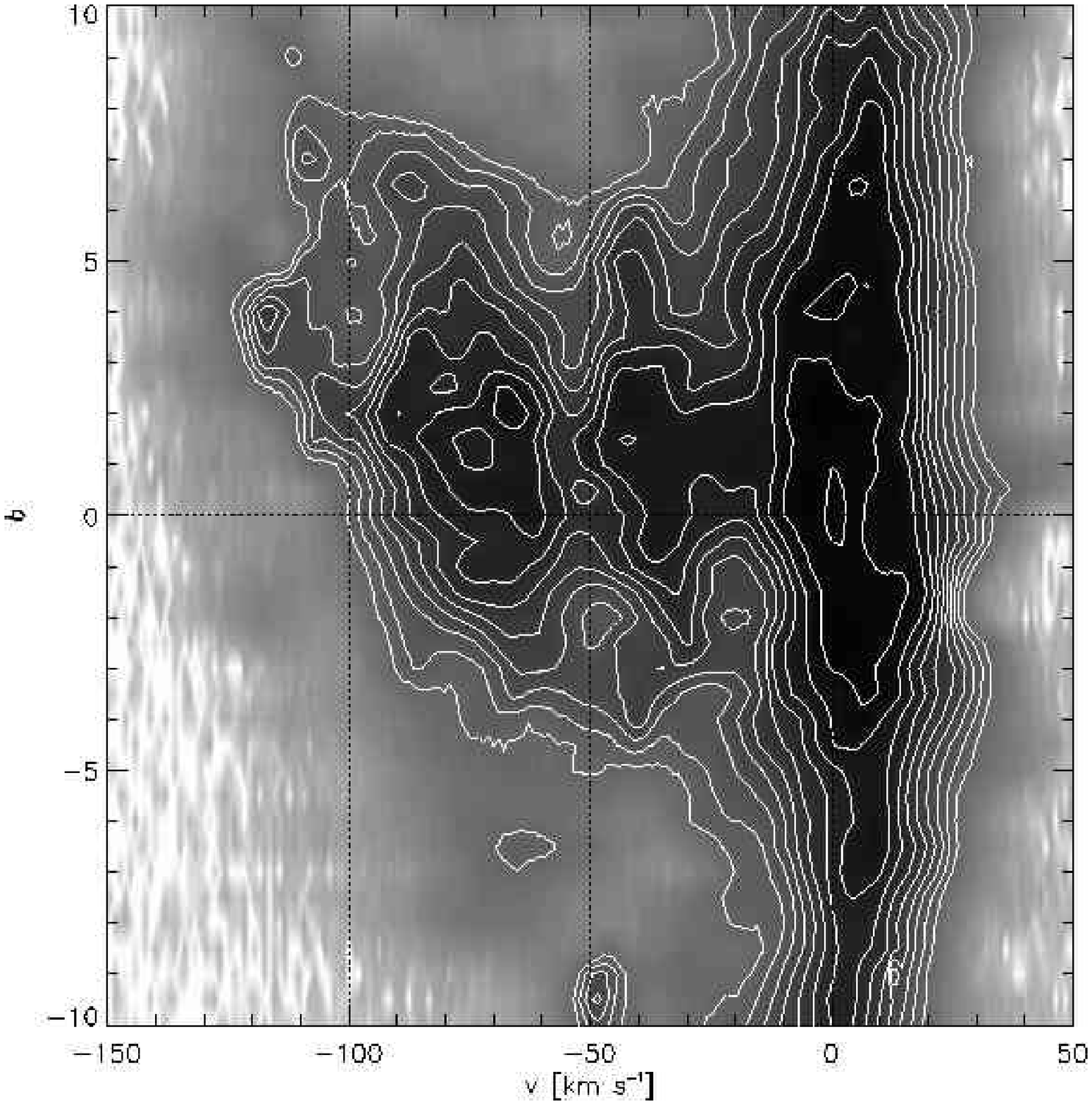}
\caption{
$v-b$ diagrams for $l=79 \grad$ for the simulation
and the Leiden\-/Dwin\-geloo survey.
The grayscale presents the column density at a given
velocity (arbitrary units for the survey) in a linear scale.
Notice that the galactic warp is present in the survey data, while
the symmetry in the simulation does not allow a warped disk.
Once again, the encounter with the edge of the calculation grid
truncates the model distribution for $v<-70 \kms$.
Note that, in both the model and the data, the stem structure
of the Norma arm (between $-40$ and $-65 \kms$) is much less
pronounced than that of the Perseus arm (between $0$ and $-35\kms$).
Its existence relies on the production of a thin dense region
ahead of the arm by the interarm downflow.
As seen in Figure \ref{dist_lat_75}, at $d \sim 8.5 \kpc$,
this region is thin, but not so dense as that ahead of the
Perseus arm.
}
\label{latvel_off}
\end{figure*}

In Figure \ref{latvel_off}, the $v-b$ diagram for the simulation
in the $l=79 \grad$ direction is compared with the equivalent diagram
for the Leiden/Dwingeloo \ion{H}{1} survey \citep{har97}.
The now familiar mushroom structures appear again at the
approximate velocities of the spiral arms,
along with the characteristic gap between caps of successive
arms.
At other galactic longitudes, the observed pattern is less
regular, presumably due to galactic complexities not found
in our model.
Notice that, in the survey data, the galactic warp displaces the
mushroom structures off the $b=0 \grad$ plane, while restrictions in
the simulation does not allow such symmetry break.


\begin{figure}[hb]
\plotone{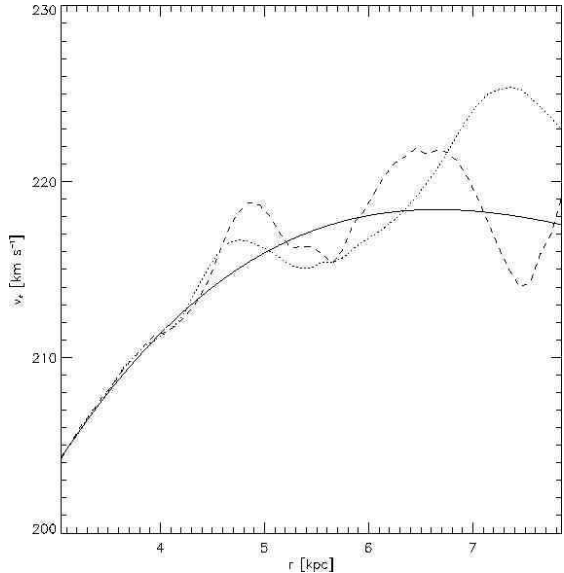}
\caption{
Rotation curve for the midplane derived by
simulating measurements in the model.
The continuous line is the rotation law in the setup, the dotted
line is the derived rotation for positive longitudes, and
the dashed line is for negative longitudes.
Our ``measured'' rotation curve is larger in the $6.8\kpc <r<8\kpc$
range for positive longitudes than for negative longitudes.
The opposite is true for the $4.3 \kpc <r <6.8 \kpc$ range.
Both trends approximate features of the measured rotation curves of
the Milky Way discussed in the text.
}
\label{rotation_curve}
\end{figure}

\section{Rotation curve.}\label{rotcurve_sec}

In order to estimate the influence of the spiral arms
on measurements of the galactic rotation curve,
we measured it in the simulation
in a way that emulates how it is measured in the Milky Way.
For gas inside the solar circle, we picked a galactic longitude.
We then looked for the maximum velocity along the line-of-sight
(minimum for negative longitudes) and assumed that it arises
from the tangent point and, therefore, the gas at that
galactocentric radius moves with that extreme velocity.
By repeating this procedure for an array of galactic
longitudes, we can trace the rotation curve
interior to the solar circle.
Figure \ref{rotation_curve} shows the results.
The dotted line shows the case for the northern
galaxy, while the dashed line shows the rotation curve for the
southern galaxy.
For comparison, the rotation curve that arises from the
hydrostatic plus rotational equilibrium in the initial conditions
is presented as the continuous line.







The observed rotation curve
\citep[when scaled for $r_\odot=8.0 \kpc$]{bli91,mcc04}
is systematically
higher in the range $55 \grad < l < 80 \grad$
($6.5 \kpc < r < 7.8 \kpc$)
than in the corresponding negative longitudes by some $7 \kms$.
The converse happens in the $40\grad < l < 55\grad$ range
($5 \kpc < r < 6.5 \kpc$).
This behavior and the amplitude of the oscillations are
reproduced by our simulations, although around $r=7.5 \kpc$,
the difference between our rotation curves, $\sim 10 \kms$, is
somewhat larger than observed.


\section{Kinematic distances.}\label{distances_sec}

An important use of the rotation curve is the estimation
of distances in the Galaxy, by assuming that the target
moves in a circular orbit.
In this section we try to estimate the error in those
distances.
The fact that the calculated curve falls below the rotation
set by the rotational hydrostatics and the background potential
(Figure \ref{rotation_curve})
creates lines of sight in which the gas never
reaches the velocity that direction should have in circular orbit,
and therefore, if this ``true rotation'' is used to estimate
distances, no gas would be assigned to those regions.
If we imagine the galaxy as made of rubber and being distorted
by the distance errors, the picture obtained
would have large holes in those regions.
For this reason, and for self-consistency,
we used the rotation curve derived in Section \ref{rotcurve_sec}.

\begin{figure}[b]
\plotone{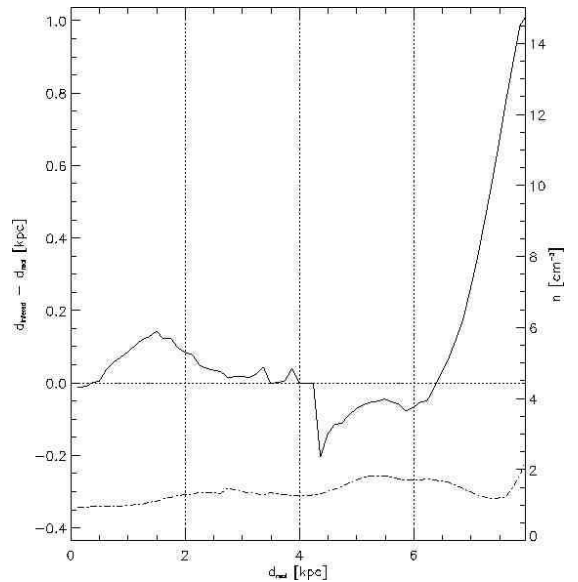}
\caption{
Error in the estimated distance due to the assumption
of circular orbits, along the $l=60\grad$ direction.
Our ``measured'' rotation curve from Figure \ref{rotation_curve}
was used to obtain the kinematic distances.
The dashed-dotted line is the gas density at that (real) distance.
Generally speaking, the distance is overestimated at the arms,
and underestimated in the interarms, although this depends on
the chosen galactic longitude.
}
\label{dist_err}
\end{figure}

\begin{figure*}
\plottwo{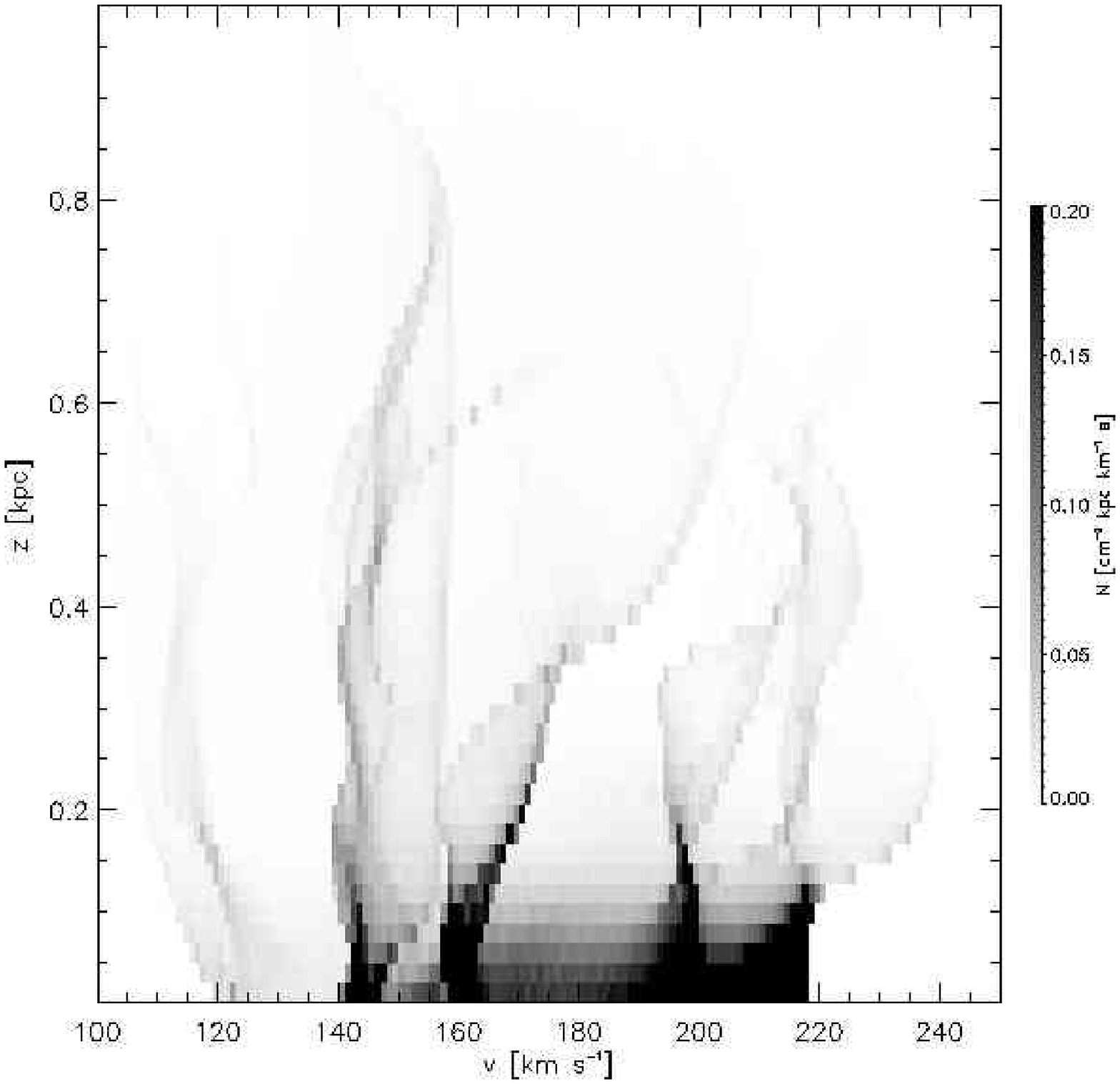}
        {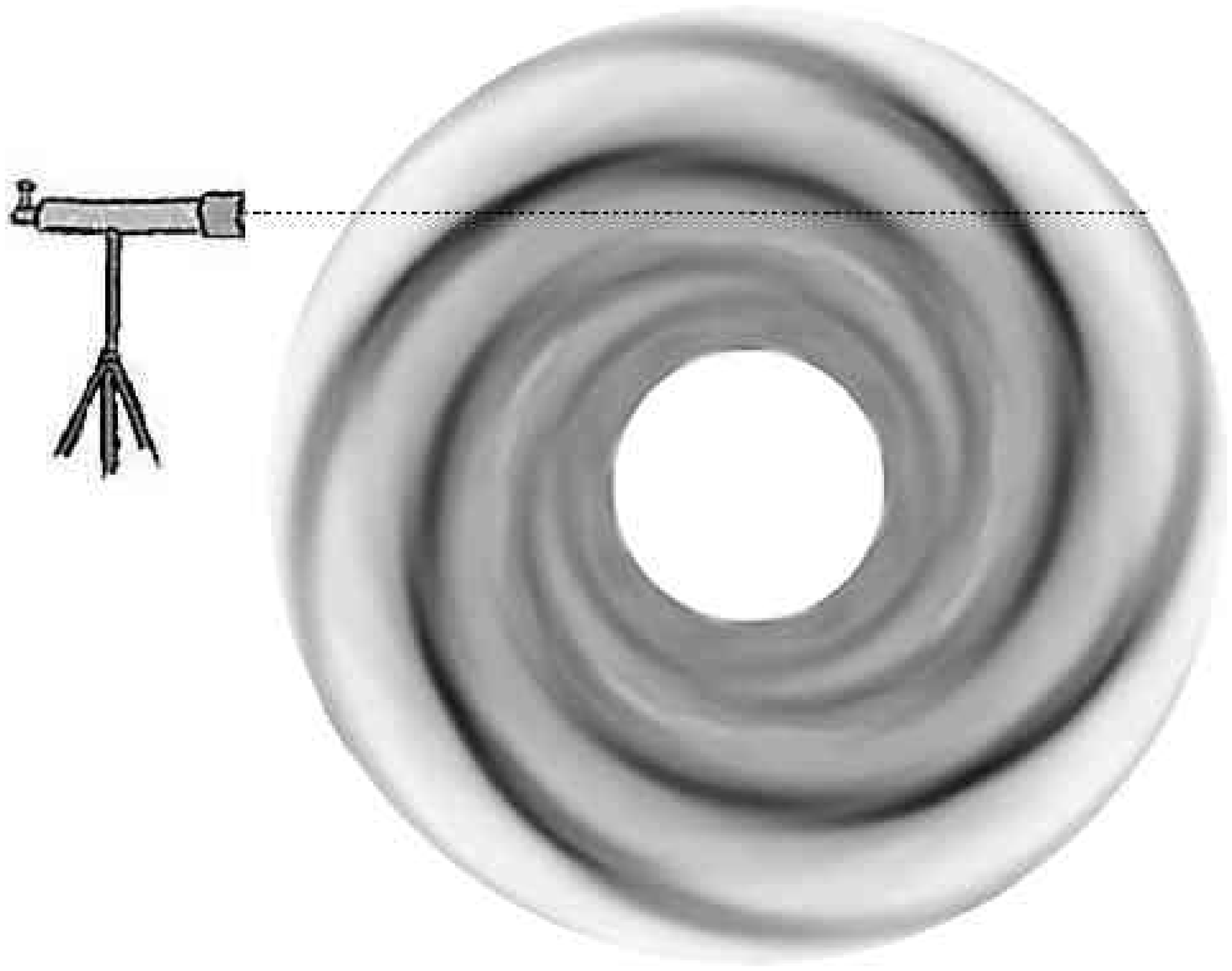}
\caption{
The left panel shows the spectrum that an observer would get
when placing the slit perpendicular to the galactic plane, at
a distance of $6.2 \kpc$ from the center.
The grayscale shows the column density for a given
velocity above the galactic plane.
As shown in the right panel, the line of sight crosses three
spiral arms, respectively from left to right,
at $160, 195$ and $145 \kms$.
The fourth midplane concentration, between 200 and $218\kms$
is along the tangent line, in an interarm region.
}
\label{corota}
\end{figure*}

Since we have the privileged information of where the
gas really is, we can estimate the error
created by assuming that the gas moves in circular orbits
with the adopted rotation curve.
At a given distance along a fixed longitude direction,
we take the velocity of the gas given by the simulation, and
calculate at what distance the circular orbit assumption puts it.
Even if the rotation curve has oscillations, the
line-of-sight velocity of the gas in circular orbits might remain
monotonic on either side of the tangent point, provided that
the oscillations are not too large
(as can be seen in Figure \ref{dist_lat_75}, that is not the case
above the plane, but here we focus in the distances along the midplane).
Therefore, there is still only a two-point ambiguity in the
assigned distance.
We solve that ambiguity by cheating: we place the gas parcel
on the side of the tangent point we know it to be.

Figure \ref{dist_err} shows the error in the estimated
distance along the $l=60\grad$ direction.
In order to compare these errors to the spiral arms, the density
is plotted also (dashed-dotted line).
Although it depends on the galactic longitude,
the distances tend to be over-estimated around the arms
and under-estimated in the interarms by a varying amount,
typically smaller than $1 \kpc$.
Gas moving at forbidden velocities cannot be given a location
with this method.


\section{Cylindrical rotation.}\label{corotation_sec}

Kinematics of the gas above the galactic plane has been
studied as a test for models for the galactic fountain, which
in turn are used to explain the ionization of the ISM at
high $z$ (see discussion in Miller \& Veilleux 2003).
Unfortunately, the velocity resolution achieved for external
galaxies makes it difficult
to estimate rotation velocity gradients over the vertical distance
our models span, and therefore comparison is difficult.
Nevertheless, we present in Figure \ref{corota} a synthetic spectrum
that could be obtained by placing the slit perpendicular to the
galactic plane, at a distance of $6.2\kpc$ from the galactic center.
Three spiral arm crossings are clear in this picture, at
$160, 195$ and $145 \kms$ (respectively, as we move away from the
observer).
Notice that the arms show a leaning toward higher velocity
as we move up, due to the gas speeding up over them after
the shock.
Also, the arm at $195\kms$ spreads over a range of velocities,
since we look at it at a smaller angle.

The right limit of the emission in this diagram is usually
associated with the rotation of the galaxy.
That maximum velocity is not reached at the point of smallest
radius, the tangent point,
but the offset is quite small ($1 \kpc$ or less).
That point corresponds to an interarm region, where the gas
falls back down to the plane, and
slows down as it moves out in radius.
Therefore, near the midplane, this maximum velocity falls
below the rotational velocity that we get when repeating this
exercise at a different radius.

As in previous cases, not all the ridges in this plot
correspond to density structures, but are generated by velocity
crowding.
Such is the case of the element at $v_{los}> 220 \kms$
around $z=200 \pc$, which is generated by a plateau of nearly
constant velocity in a region with density lower than its
surroundings, which have a larger spread in velocity.

From Figure \ref{corota}, we would infer that the velocity
maximum increases
some $20\kms$ up to $z=200\pc$, and then comes back down
to $220 \kms$ around $z=400\pc$.
When placing the slit at a different position, we get similar
behavior, with different amplitudes and at different heights.
\citet{mil03} observed a similar behavior in a couple of galaxies
in their sample, although with a larger velocity amplitude.
Those seem to be exceptions, since most of the galaxies in their
sample do not show significant gradients.
The amplitude of the velocity gradient in this model would
have been difficult to detect with the resolution
achieved in their study.


\section{Synchrotron emission.}\label{synch_sec}

\begin{figure}[t]
\plotone{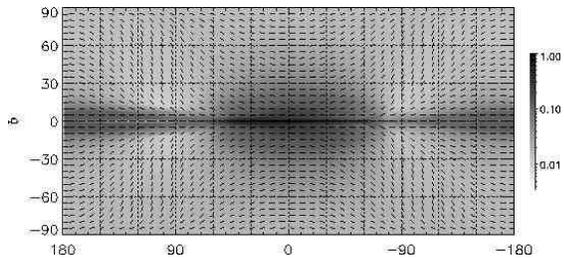}
\caption{
All sky map of synchrotron emission, in arbitrary units.
The dashes mark the direction of the magnetic field, as implied
from the polarized emission.
With exception of the midplane, a circular field pattern
is observed, although the field in the simulation is not.
}
\label{sync_lon}
\end{figure}

Figure \ref{sync_lon} shows an all sky map of the synchrotron
emission.
The grayscale corresponds to the total synchrotron intensity,
while the direction of the magnetic field,
as inferred from the
polarization of the integrated emission, is presented in dashes.
The synchrotron emissivity is given by

\begin{equation}
\varepsilon_{tot}(r,\phi) \propto n_{cr}(r,z) B_\perp^{(p+1)/2},
\end{equation}

\noindent
where $B_\perp$ is the component of the magnetic field
perpendicular to the line-of-sight,
$p=2.5$ is the spectral index of the distribution of
cosmic ray electrons,
$n_{cr}=\exp(-r/r_{cr}-z/z_{cr})$ is its space density,
$r_{cr}=13 \kpc$ and $z_{cr}=2.5 \kpc$ \citep{fer98}.
For each $(l,b)$ direction, this emissivity is integrated,
and the direction of the polarization is accounted for by
the process described in Paper II.

The degree of polarization of the integrated emission is
generally high, varying
from $>70\%$ in the midplane and in the $l\sim 0\grad$ and
$180\grad$ at all latitudes, to $\sim 40\%$ in four isolated
regions toward $(l,b) \sim (\pm 120\grad, \pm 70\grad)$.
This is expected, since our resolution does not allow us to
model the random component of the field.

\begin{figure}[b]
\plotone{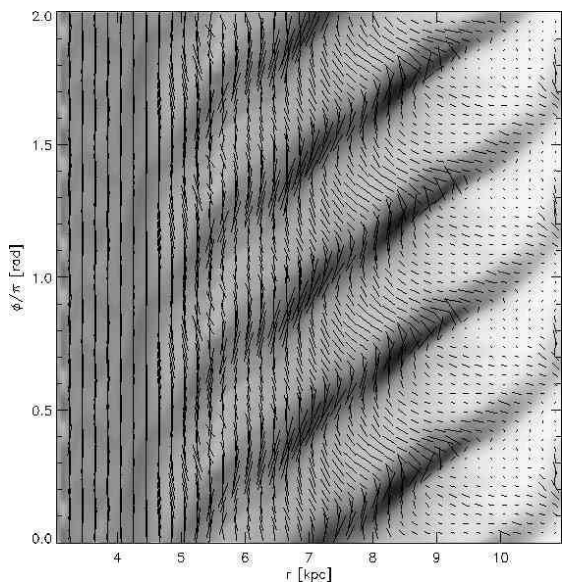}
\caption{
Reproduced from Paper II, this Figure presents the
direction of the magnetic field in the midplane, displayed in
a polar diagram.
The axes show the radius from the galactic center and the azimuthal
angle (the Sun would be in the $(r,\phi) = (8 \kpc, \pi/2)$ position).
The orbiting gas flows down from the top.
The length of the dashes is proportional to the field strength.
The grayscale represents the surface density of the
simulation.
Notice that the field adopts a negative pitch angle in the
interarm region.
}
\label{mag_pitch}
\end{figure}

To help with the discussion of this map, we present
the midplane magnetic field in the simulation
in Figure \ref{mag_pitch}, reproduced from Paper II.
The grayscale represents the surface density, and
the dashes follow the direction of the magnetic field, with
its length proportional to the field strength.

Recalling Figure \ref{arms}, the second and third quadrants
have interarm gas nearby.
The Perseus arm is $2 \kpc$ away in the $l=180\grad$ direction,
and $5 \kpc$ away toward $l=90\grad$.
In that $90\grad < l < 180\grad$ range, near the midplane,
the magnetic field has a negative pitch angle.
This has the effect that the field is pointing directly toward
the observer, and the plane-of-the-sky component is very small,
causing the low emission at low latitudes in Figure \ref{sync_lon}.
Around the $l=-90\grad$ direction, that negative pitch angle
increases the component of the field projected in the sky,
causing a higher emission in the $-90\grad < l < -60\grad$
range.
On the other hand, the field above the plane twists more abruptly,
with a smaller negative pitch region,
yielding a short path length to integrate over and leaving
only a small imprint on this plot.
As a consequence, off the galactic plane, we see a pattern
in the sky that resembles that of a circular field:
horizontal in the $l\sim 0\grad$ and
$180\grad$ at all latitudes, and in the midplane, at all
longitudes; nearly vertical off the plane
in the $l \sim 90\grad$ and $-90\grad$ directions
(the sky projection of an overhead circle).

Although the sign of the radial component of the field changes
between the arm and interarm regions,
those changes are not enough to
account for the changes in the differential rotation measure that
are typically interpreted as reversals in the direction of the
galactic magnetic field \citep[and references therein]{bec01}.


\section{Conclusions.}\label{conclusions_sec}

By placing an imaginary observer inside the modeled galaxy
from Paper II, we generated various synthetic observations.
Since we actually know where the gas is, we can distinguish which
parts of the model are generating the observed structures.
In particular, velocity crowding effects can be distinguished
from real spatial concentrations.

The synthetic $l-v$ diagram has common characteristics with
the observed diagram, although some similarities are only
in a qualitative level.
Again, velocity crowding generates ridges that do not correspond
to spiral arms.
But the converse also happens, as a large velocity gradient can
dilute a spiral arm in velocity space.
In the $v-b$ diagram, on the other hand, velocity crowding and
dilution generate structures that test the
vertical distribution of matter and velocity above the arms
presented in Papers I and II.
Such structures can be observed in the Leiden/Dwingeloo \ion{H}{1}
survey.

We also explored the rotation curve that the imaginary observer
would measure from within the model galaxy.
Several of the characteristics of this measured rotation curve
are also observed in the curve measured for the Milky Way,
being higher in some radial ranges
than on the opposite side of the galactic center.
This agreement is probably a consequence
of our having tried to fit
the positions of the spiral arms in our model to
the proposed positions for the Galaxy.
Nevertheless, the measured rotation curve has large deviations
with respect to the rotation from our initial conditions,
which is also influenced by pressure gradients and magnetic
tension and is therefore slightly different
from the rotation consistent
with the background gravitational potential.

Although the magnetic field is largely non-circular, the
averaging effect of the synthetic synchrotron maps generates
a largely circular imprint, with exception of the midplane.
As mentioned in Paper II, restrictions in our model limit the
amount of vertical field that our model generates,
increasing the circularity of the field and diminishing
the vertical extent of the synchrotron emission.

The ISM of the Milky Way Galaxy is very complex system, in which
many different physical processes combine to generate
large scale structure.
The extra freedom of the third dimension and the dynamical
effects of a strong magnetic field, in our view, are two
key elements in the formation of such structures.
Until we find a way of reliably measuring distances to the diffuse
components of the disk, velocity crowding effects will keep
blurring and distorting the pictures we generate.
More realistic modeling with higher resolution, inclusion
of gas self-gravity, and stellar feedback
(including cosmic ray generation and diffusion) is necessary to
further clarify that picture.


\acknowledgements

We thank R. Benjamin, E. Wilcots, N. McClure-Griffiths,
G. Madsen, and J. Lockman for useful
comments and suggestions, to the NASA Astrophysics Theory
Program for financial support under the grant NAG 5-12128,
and to M\'exico's Consejo Nacional de Ciencia y Tecnolog\'{\i}a
for support to G. C. G.


{}


\begin{thebibliography}{}

\bibitem[Beck(2001)]{bec01}
  Beck, R. 2001, Space Sci. Rev., 99, 243

\bibitem[Blitz \& Spergel(1991)]{bli91}
  Blitz, L. and Spergel, D. N. 1991, \apj, 370, 205

\bibitem[Burton(1971)]{bur71}
  Burton, W. B. 1971, \aap, 10, 76

\bibitem[Dieter(1964)]{die64}
  Dieter, N. H. 1964, \aj, 69, 137

\bibitem[Ferri\`ere(1998)]{fer98} Ferri\`ere, K.
  1998, \apj, 497, 759

\bibitem[Georgelin \& Georgelin(1976)]{geo76}
  Georgelin, Y. M. and Georgelin, Y. P. 1976, \aap, 49, 57

\bibitem[G\'omez \& Cox(2002)]{gom02}
  G\'omez, G. C. and Cox, D. P. 2002, \apj, 580, 235 (Paper I)

\bibitem[G\'omez \& Cox(2004)]{gom04}
  G\'omez, G. C. and Cox, D. P. 2004, \apj, {\it submitted}
  (Paper II)

\bibitem[Haffner et al.(2003)]{haf03}
  Haffner, L. M., Reynolds, R. J., Tufte, S. L., Madsen, G. J.,
  Jaehnig, K. P. and Percival, J. W. 2003,
  \apjs, 149, 405

\bibitem[Hartmann \& Burton(1997)]{har97}
  Hartmann, D. and Burton, W. B. 1997, Atlas of Galactic
  Neutral Hydrogen (Cambridge: Cambridge Univ. Press)

\bibitem[McClure-Griffiths et al.(2004)]{mcc04}
  McClure-Griffiths, N. M., Dickey, J. M., Gaensler, B. M.
  and Green, A. J. 2004, \apj, {\it submitted}

\bibitem[Miller \& Veilleux(2003)]{mil03}
  Miller, S. T. and Veilleux, S. 2003, \apj, 592, 79

\bibitem[Mulder \& Liem(1986)]{mul86}
  Mulder, W. A. and Liem, B. T. 1986, \aap, 157, 148

\bibitem[Taylor \& Cordes(1993)]{tay93}
  Taylor, J. H. and Cordes, J. M. 1993, \apj, 411, 674

\end{thebibliography}
\end{document}